\documentclass[twocolumn,showpacs,showkeys,eqsecnum,superscriptaddress,twoside,bibnotes]{revtex4}

\usepackage{amsmath}
\usepackage{amssymb}
\usepackage{graphicx}

\begin{document}

\title{Anderson Localization in Metamaterials with Compositional Disorder}

\author{E.~J.~Torres-Herrera}
\affiliation{Instituto de F\'{\i}sica, Universidad Aut\'{o}noma de Puebla, Apartado Postal J-48, Puebla, Puebla, 72570, Mexico}

\author{F.~M.~Izrailev}
\affiliation{Instituto de F\'{\i}sica, Universidad Aut\'{o}noma de Puebla, Apartado Postal J-48, Puebla, Puebla, 72570, Mexico}

\author{N.~M.~Makarov}
\email{makarov.n@gmail.com}
\affiliation{Instituto de Ciencias, Universidad Aut\'{o}noma de Puebla, Priv. 17 Norte No 3417, Col. San Miguel Hueyotlipan, Puebla, Puebla, 72050, Mexico}

\date{\today}

\begin{abstract}
We consider one-dimensional periodic-on-average bi-layered models with random perturbations in dielectric constants of both basic slabs composing the structure unit-cell. We show that when the thicknesses $d_a$ and $d_b$ of basic layers are essentially nonequal, $d_a\neq d_b$, the localization length $L_{loc}$ is described by the universal expression for two cases: (a) both layers are made from right-handed materials (the RH-RH model), (b) the $a$ layers are of a right-handed material while the $b$ layers are of a left-handed material (the RH-LH model). For these models the derived expression for $L_{loc}$ includes all possible correlations between two disorders. However, when $d_a=d_b$ the RH-LH model exhibits a highly non-trivial properties originated from inhomogeneous distribution of the phase of propagating wave, even in the case of white-noise disorder. We analytically show that in this case the localization length diverges in the conventional second order in perturbation parameters. Therefore, recently numerically discovered anomalies in $L_{loc}$ are due to the next order of approximation. On the other hand, for the RH-RH model the general expression for $L_{loc}$ remains valid for $d_a=d_b$ as well.
\end{abstract}


\maketitle

\section{Introduction}

Due to progress in nano- and material science, the study of wave propagation and electron transport in periodic one-dimensional (1D) systems has attracted much attention (see, e.g., Ref.~\onlinecite{MS08} and references therein). The systems of particular interest are bi-layer structures in optics and electromagnetics, or semiconductor superlattices and arrays of alternating quantum wells and barriers in electronics. The interest to this subject is due to a possibility to create structures with prescribed properties of transmission and/or reflection. New perspectives in this direction are related to unconventional optic properties of metamaterials.

One of the practical problems is the influence of disorder that cannot be avoided in real experiments. The disorder can be originated from fluctuations of the thickness of layers (\emph{positional disorder}) or from variations of the medium parameters, such as permittivity and permeability for electromagnetic waves or the barrier height for electrons (\emph{compositional disorder}). Typically, the disorder is treated as an unwanted effect, however, recently it was understood that it can be a promising candidate for targeted manipulation of transport properties. Indeed, the correlations in the disorder may result in unusual features of transport. In particular, it was shown, both theoretically \cite{IK99,KI99,IKU01,IM01,IM03,IM04,IM05} and experimentally \cite{KIKS00,KIK08,DKSMI11}, that specific long-range correlations can significantly enhance or suppress the wave localization in a desired window of frequency.

As is well known, the transmission through any 1D disordered system is governed by Anderson localization (see, e.g., Refs.~\onlinecite{IM05,LGP88,M99} and references therein). The principal concept of this phenomenon is that all transport characteristics depend solely on the ratio $L/L_{loc}$ between the structure length $L$ and \emph{localization length} $L_{loc}$ of eigenstates. Such a universal dependence manifests itself, for example, in the famous expression for the self-averaging logarithm of the transmittance $T_L$,
\begin{equation}\label{FNT-AvLn}
\langle\ln T_L\rangle=-2L/L_{loc}.
\end{equation}
Here, the angular brackets $\langle\ldots\rangle$ stand for averaging over the disorder. In agreement with the concept of single parameter scaling, there are only two characteristic regimes in 1D disordered structures, namely, the regimes of ballistic and localized transport. The ballistic transport occurs when the localization length $L_{loc}$ is much larger than the sample length $L$. In this case the samples are practically fully transparent since its average transmittance is close to unity,
\begin{equation}\label{FNT-Tbal}
\langle T_L\rangle\approx1-2L/L_{loc}\qquad\mbox{for}\qquad L\ll L_{loc}.
\end{equation}
In the opposite case when $L_{loc}$ is much smaller than $L$, the 1D disordered structures exhibit localized transport. In this case the average transmittance is exponentially small,
\begin{equation}\label{FNT-repavT}
\langle T_L\rangle\approx\exp(-2L/L_{loc})\qquad\mbox{for}\quad L_{loc}\ll L.
\end{equation}
As one can see, in the localization regime a 1D disordered system perfectly (with an exponential accuracy) reflects quantum or classical waves because of strong localization of all eigenstates. This brief analysis shows how one can control the transport by manipulating the value of localization length in comparison with the system size.

In spite of a remarkable progress in understanding the main features of the wave (electron) propagation through random structures, the majority of studies are based on numerical methods \cite{MCMM93,So00,SSS01,Po03,VM03,Eo06,DZ06,No07,Po07,Ao07,NML07,NML08}.
Giving the important results the numerical approaches obviously suffer from the lack of generality being restricted by specific values of parameters. As for the analytical results, they are mainly refer to simplest models with white-noise disorder \cite{BW85,LIMKS09,LM09,Ao10,Mo10} or to the correlated disorder with {\it delta-like} potentials in the Anderson \cite{IK99} or Kronig-Penney models \cite{IKU01,HIT08}.

In this paper we address the 1D models with periodic-on-average bi-layered structures with weakly disordered parameters. In comparison with Refs.~\onlinecite{MIL07,IM09} where the case of random thicknesses of slabs was considered, here we focus on the model with weakly disordered refractive indices of both slabs. The main attention is paid to the comparison between normal stacks (when both layers are of normal material) and mixed stacks (with alternating normal and meta-material layers). First, we review the recent study of the structures with different thicknesses of two basic layers \cite{IMT10,MIT10}. Here we derive the unique analytical expression for the localization length $L_{loc}$, which is valid in a large range of model parameters, and can be applied to various physical realizations. The key point of our approach is that we explicitly take into account possible correlations within two disorders of each layer type as well as between them. Second, we extend our analytical analysis to the particular case of equal thicknesses of basic layers, see Ref.~\onlinecite{Ao07}. We show that in this specific case the standard perturbation theory fails if one of two basic stacks is made from a left-handed material. Specifically, the localization length diverges in the conventional second order in perturbation parameters, thus leading to anomalous properties of scattering.

\section{Problem formulation}
\label{PF}

 The model describes the propagation of electromagnetic waves of frequency $\omega$ along an infinite array of two alternating $a$ and $b$ layers. Every kind of slabs is respectively specified by the dielectric permittivity $\varepsilon_{a,b}$, magnetic permeability $\mu_{a,b}$, corresponding refractive index $n_{a,b}=\sqrt{\varepsilon_{a,b}\mu_{a,b}}$, impedance $Z_{a,b}=\mu_{a,b}/n_{a,b}$ and wave number $k_{a,b}=\omega n_{a,b}/c$. We consieder two cases: (a) when both $a$ and $b$ slabs consist of a right-handed (RH) optic materials, and (b) when $a$ slabs are of RH-material and $b$ layers are of left-handed (LH) material. In what follows, the combination of RH-RH slabs is referred to the \emph{homogeneous} stack whereas the array of RH-LH layers is called \emph{mixed} stack.

As is known, for the RH medium all optic parameters are positive. On the contrary, for the LH material the permittivity, permeability and corresponding refractive index are negative, however the impedance remains positive. Every alternating slab has the constant thickness $d_a$ or $d_b$, respectively, so that the size $d$ of the unit $(a,b)$ cell is also constant, $d=d_a+d_b$.

As was noted in Ref.~\onlinecite{IM09}, when the impedances of two basic $a$ and $b$ slabs are matched, $Z_a=Z_b$, the localization length diverges and the perfect transparency emerges, while a \emph{positional} disorder itself persists. In the following, we analyze  the effect of \emph{compositional disorder} in a stack-structure whose unperturbed counterpart consists of perfectly matched basic $a$ and $b$ layers. Specifically, following Ref.~\onlinecite{Ao07}, we admit that a disorder is incorporated via the dielectric constants $\varepsilon_{a,b}$ only, so that
\begin{subequations}\label{FNT-nZanZb}
\begin{eqnarray}
&&\varepsilon_a(n)=[1+\eta_a(n)]^2,\quad\mu_a=1,\nonumber\\[6pt]
&&n_a(n)=1+\eta_a(n),\quad Z_a(n)=[1+\eta_a(n)]^{-1};\label{FNT-nZa}\\[6pt]
&&\varepsilon_b(n)=\pm[1+\eta_b(n)]^2,\quad\mu_b=\pm1,\nonumber\\[6pt]
&&n_b(n)=\pm[1+\eta_b(n)],\quad Z_b(n)=[1+\eta_b(n)]^{-1}.\qquad\label{FNT-nZb}
\end{eqnarray}
\end{subequations}
Here integer $n$ enumerates the unit $(a,b)$ cells. The upper sign stands for the RH material while the lower one is associated with LH medium.

Without disorder, $\eta_{a,b}(n)=0$, the homogeneous RH-RH structure is just the air without any stratification, while the mixed RH-LH array represents the so-called \emph{ideal mixed stack} ($\varepsilon_a=\mu_a=1$, $\varepsilon_b=\mu_b=-1$) with perfectly matched slabs ($Z_a=Z_b=1$). Therefore, both the unperturbed RH-RH and RH-LH stacks are equivalent to the homogeneous media with the refractive index $\overline{n}$, thus resulting in no gaps in their linear spectrum,
\begin{equation}\label{FNT-UnpDR}
\kappa=\omega\overline{n}/c,\qquad\overline{n}=\frac{|d_a\pm d_b|}{d_a+d_b}.
\end{equation}

The random sequences $\eta_{a}(n)$ and $\eta_{b}(n)$ describing the compositional disorder, are statistically homogeneous with zero average, $\langle\eta_{a,b}(n)\rangle=0$, and binary correlation functions defined by
\begin{subequations}\label{FNT-CorrDef}
\begin{eqnarray}
&&\langle\eta_a(n)\eta_a(n')\rangle=\sigma_a^2K_a(n-n')\,,\\[6pt]
&&\langle\eta_b(n)\eta_b(n')\rangle=\sigma_b^2K_b(n-n')\,,\\[6pt]
&&\langle\eta_a(n)\eta_b(n')\rangle=\sigma_{ab}^2K_{ab}(n-n')\,.
\end{eqnarray}
\end{subequations}
The averaging $\langle ... \rangle$ is performed over the whole array or due to the ensemble averaging, that is assumed to be equivalent. The auto-correlators $K_{a}(r)$ and $K_{b}(r)$ as well as the cross-correlator $K_{ab}(r)$ are normalized to unity, $K_{a}(0)=K_{b}(0)=K_{ab}(0)=1$, and decrease with an increase of the distance $|r|=|n-n'|$ between the cell indices $n$ and $n'$. The variances $\sigma^2_{a}$ and $\sigma^2_{b}$ are obviously positive, however, the term $\sigma_{ab}^2$ can be of arbitrary value (positive, negative or zero). We assume the compositional disorder to be weak, i.e.
\begin{equation}\label{FNT-WD}
\sigma_{a,b}^2\ll1,\qquad(k_{a,b}d_{a,b})^2\sigma_{a,b}^2\ll1;
\end{equation}
that allows us to develop a proper perturbation theory. In this case all transport properties are entirely determined by the randomness power spectra $\mathcal{K}_a(k)$, $\mathcal{K}_b(k)$, and $\mathcal{K}_{ab}(k)$, defined by the standard relations
\begin{subequations}\label{FNT-FT-K}
\begin{eqnarray}
\mathcal{K}(k)&=&\sum_{r=-\infty}^{\infty}K(r)\exp(-ikr),\\[6pt]
K(r)&=&\frac{1}{2\pi}\int_{-\pi}^{\pi}dk\,\mathcal{K}(k)\exp(ikr).
\end{eqnarray}
\end{subequations}
By definition \eqref{FNT-CorrDef}, all the correlators $K_{a}(r)$, $K_{b}(r)$ and $K_{ab}(r)$ are real and even functions of the difference $r=n-n'$ between cell indices. Therefore, the corresponding Fourier transforms $\mathcal{K}_a(k)$, $\mathcal{K}_b(k)$ and $\mathcal{K}_{ab}(k)$ are real and even functions of the dimensionless wave number $k$. It should be also stressed that according to rigorous mathematical theorem, the power spectrum $\mathcal{K}(k)$ is non-negative for any real sequence.

Within any of $a$ or $b$ layers the electric field of the wave $\psi(x)\exp(-i\omega t)$ obeys the 1D Helmholtz equation with two boundary conditions at the interfaces between neighboring slabs,
\begin{subequations}\label{FNT-WaveEqBC}
\begin{eqnarray}
&&\left(\frac{d^2}{dx^2}+k_{a,b}^2\right)\psi_{a,b}(x)=0,\label{FNT-WaveEq-ab}\\[6pt]
&&\psi_a(x_i)=\psi_b(x_i),\qquad \mu_a^{-1}\psi'_a(x_i)=\mu_b^{-1}\psi'_b(x_i).\qquad\label{FNT-BC}
\end{eqnarray}
\end{subequations}
The $x$-axis is directed along the array of bi-layers perpendicular to the stratification, with $x=x_{i}$ standing for the interface coordinate.

\section{Basic expressions}
\label{BE}

The general solution of Eq.~\eqref{FNT-WaveEq-ab} within the $n$th $(a,b)$ cell can be presented in the following form,
\begin{subequations}\label{FNT-Psi-ab}
\begin{eqnarray}
\psi_{a}(x)&=&\psi_{a}(x_{an})\cos\left[k_a(x-x_{an})\right]\nonumber\\[6pt]
&&+k_a^{-1}\psi'_a(x_{an})\sin\left[k_a(x-x_{an})\right]\\[6pt]
\mbox{inside}\,&a_n&\mbox{layer, where}\,x_{an}\leq x\leq x_{bn}\,;\nonumber\\[6pt]
\psi_{b}(x)&=&\psi_{b}(x_{bn})\cos\left[k_b(x-x_{bn})\right]\nonumber\\[6pt]
&&+k_b^{-1}\psi'_b(x_{bn})\sin\left[k_b(x-x_{bn})\right]\\[6pt]
\mbox{inside}\,&b_n&\mbox{layer, where}\,x_{bn}\leq x\leq x_{a(n+1)}\,.\nonumber
\end{eqnarray}
\end{subequations}
The coordinates $x_{an}$ and $x_{bn}$ refer to the left-hand edges of successive $a_n$ and $b_n$ layers. Note that $x_{bn}-x_{an}=d_a$ and $x_{a(n+1)}-x_{bn}=d_b$. The solution \eqref{FNT-Psi-ab} gives a useful relation between the wave function $\psi_{a,b}$ and its derivative $\psi'_{a,b}$ at the opposite boundaries of the same $a$ or $b$ layer,
\begin{subequations}\label{FNT-Map-ab}
\begin{eqnarray}
\psi_a(x_{bn})&=&\psi_a(x_{an})\cos\widetilde{\varphi}_a(n)\nonumber\\
&&+k_a^{-1}\psi'_a(x_{an})\sin\widetilde{\varphi}_a(n),\nonumber\\[6pt]
\psi'_a(x_{bn})&=&-k_a\psi_a(x_{an})\sin\widetilde{\varphi}_a(n)\nonumber\\
&&+\psi'_a(x_{an})\cos\widetilde{\varphi}_a(n);\\[6pt]
\psi_b(x_{a(n+1)})&=&\psi_b(x_{bn})\cos\widetilde{\varphi}_b(n)\nonumber\\
&&+k_b^{-1}\psi'_b(x_{bn})\sin\widetilde{\varphi}_b(n),\nonumber\\[6pt]
\psi'_b(x_{a(n+1)})&=&-k_b\psi_b(x_{bn})\sin\widetilde{\varphi}_b(n)\nonumber\\
&&+\psi'_b(x_{bn})\cos\widetilde{\varphi}_b(n).
\end{eqnarray}
\end{subequations}
The corresponding phase shifts $\widetilde{\varphi}_{a}(n)$ and $\widetilde{\varphi}_{b}(n)$ depend on the cell index $n$ via random refractive indices \eqref{FNT-nZanZb},
\begin{subequations}\label{FNT-phi-ab}
\begin{eqnarray}
\widetilde{\varphi}_{a}(n)=k_a(n)d_a=\varphi_{a}[1+\eta_{a}(n)],\quad\varphi_{a}=\omega d_a/c;\quad\\[6pt]
\widetilde{\varphi}_{b}(n)=k_b(n)d_b=\varphi_{b}[1+\eta_{b}(n)],\quad\varphi_{b}=\pm\,\omega d_b/c.\quad
\end{eqnarray}
\end{subequations}
By combining Eqs.~\eqref{FNT-Map-ab} with boundary conditions \eqref{FNT-BC} at $x_i=x_{bn}$ and $x_i=x_{a(n+1)}$, one can write the recurrent relations for two opposite edges of the whole $n$th unit $(a,b)$ cell,
\begin{equation}\label{FNT-mapQP}
Q_{n+1}=A_nQ_n+B_nP_n,\quad P_{n+1}=-C_nQ_n+D_nP_n.
\end{equation}
Here the ``coordinate" $Q_n$ and ``momentum" $P_n$ are
\begin{equation}\label{FNT-QPdef}
Q_n =\psi_{a}(x_{an})\quad\mbox{and}\quad P_n=(c/\omega)\psi'_{a}(x_{an}),
\end{equation}
with indices $n$ and $n+1$ standing for left and right edges of the $n$th cell. The factors $A_n$, $B_n$, $C_n$, $D_n$ read
\begin{subequations}\label{FNT-ABCDn}
\begin{eqnarray}
A_n&=&\cos\widetilde{\varphi}_a\cos\widetilde{\varphi}_b-Z_a^{-1}Z_b\sin\widetilde{\varphi}_a\sin\widetilde{\varphi}_b,\\
B_n&=&Z_a\sin\widetilde{\varphi}_a\cos\widetilde{\varphi}_b+Z_b\cos\widetilde{\varphi}_a\sin\widetilde{\varphi}_b,\\
C_n&=&Z_a^{-1}\sin\widetilde{\varphi}_a\cos\widetilde{\varphi}_b+Z_b^{-1}\cos\widetilde{\varphi}_a\sin\widetilde{\varphi}_b,\\
D_n&=&\cos\widetilde{\varphi}_a\cos\widetilde{\varphi}_b-Z_aZ_b^{-1}\sin\widetilde{\varphi}_a\sin\widetilde{\varphi}_b.
\end{eqnarray}
\end{subequations}
They depend on the cell number $n$ due to random refractive indices \eqref{FNT-nZanZb} that enter into both the impedances $Z_{a,b}(n)$ and phase shifts $\widetilde{\varphi}_{a,b}(n)$. It is noteworthy that the recurrent relations \eqref{FNT-mapQP} can be treated as the Hamiltonian map of trajectories in the phase space $(Q,P)$ with discrete time $n$ for a linear oscillator subjected to time-depended parametric force.

With vanishing disorder, $\eta_{a,b}(n)=0$, the factors \eqref{FNT-ABCDn} do not depend on the cell index (time) $n$. Therefore, in line with map \eqref{FNT-mapQP}, the trajectory $Q_n,P_n$ creates a circle in the phase space $(Q,P)$ that is an image of the unperturbed motion,
\begin{eqnarray}\label{FNT-mapQPunp}
Q_{n+1}=Q_n\cos\gamma+P_n\sin\gamma,\nonumber\\[6pt]
P_{n+1}=-Q_n\sin\gamma+P_n\cos\gamma.
\end{eqnarray}
The unperturbed phase shift $\gamma$ over a single unit $(a,b)$ cell is defined as
\begin{equation}\label{FNT-gamma-def}
\gamma=\varphi_a+\varphi_b=\omega(d_a\pm d_b)/c.
\end{equation}
This result is in a complete correspondence with the spectrum \eqref{FNT-UnpDR} taking into account that the Bloch wave number $\kappa=|\gamma|/d$.

Having the circle \eqref{FNT-mapQPunp}, it is suitable to pass to polar coordinates $R_n$ and $\theta_n$ via the standard transformation
\begin{equation}\label{FNT-QP-RTheta}
Q_n=R_n\cos\theta_n,\qquad P_n=R_n\sin\theta_n.
\end{equation}
By direct substitution of Eq.~\eqref{FNT-QP-RTheta} into map \eqref{FNT-mapQPunp}, one can see that for the unperturbed trajectory the radius $R_n$ is conserved, while the phase $\theta_n$ changes by the \emph{Bloch phase} $\gamma$ in one step of time $n$,
\begin{equation}\label{FNT-mapRThetaUnp}
R_{n+1}=R_n,\qquad\theta_{n+1}=\theta_n-\gamma.
\end{equation}

Evidently, the random perturbations, $\eta_{a,b}(n)\neq0$, give rise to a distortion of the circle \eqref{FNT-mapRThetaUnp} that is described by the Hamiltonian map \eqref{FNT-mapQP} with randomized factors \eqref{FNT-ABCDn}. With the use of definition \eqref{FNT-QP-RTheta}, one can readily rewrite this disordered map in the radius-angle presentation. The corresponding exact recurrent relations read
\begin{subequations}\label{FNT-mapRTheta}
\begin{eqnarray}
&&\frac{R_{n+1}^2}{R_n^2}=(A_n^2+C_n^2)\cos^2\theta_n+(B_n^2+D_n^2)\sin^2\theta_n\nonumber\\[6pt]
&&+(A_n B_n-C_n D_n)\sin2\theta_n\,,\label{FNT-mapR}\\[6pt]
&&\tan\theta_{n+1}=\frac{-C_n+D_n\tan\theta_n}{A_n+B_n\tan\theta_n}\,.\label{FNT-mapTheta}
\end{eqnarray}
\end{subequations}
Eqs.~\eqref{FNT-mapRTheta} constitute the complete set of equations in order to derive the localization length $L_{loc}$ according to its definition via the Lyapunov exponent $\lambda$ \cite{IKT95,IRT98},
\begin{equation}\label{FNT-LyapDef}
\frac{d}{L_{loc}}\equiv\lambda=\frac{1}{2}\Big\langle\ln\left(\frac{R_{n+1}}{R_n}\right)^2\Big\rangle\,.
\end{equation}
Note that in Eq.~\eqref{FNT-LyapDef} the averaging is performed along the trajectory specified by $R_n$ and $\theta_n$.

Here we should emphasize the following. In the ideal mixed stack ($\varepsilon_a=\mu_a=1$, $\varepsilon_b=\mu_b=-1$, $Z_a=Z_b=1$) the wave spectrum \eqref{FNT-UnpDR} is singular. Specifically, when thicknesses $d_a$ and $d_b$ are equal, $d_a=d_b$, the phase velocity $c/\overline{n}$ diverges and the Bloch phase $\gamma$ vanishes. As a result, the circle \eqref{FNT-mapRThetaUnp} presenting the unperturbed map, degenerates into a point in the phase space $(Q,P)$. Therefore, the perturbation theory has to be developed in a different way depending on the value, finite or vanishing, of the Bloch phase $\gamma$. For this reason, in what follows we perform the separate analysis for $d_a=d_b$ case when considering the RH-LH stack-structure.

\section{Bi-layer array with finite Bloch phase}
\label{finiteGamma}

In this section we assume the arbitrary relation between slab thicknesses, $d_a$ and $d_b$, for the homogeneous RH-RH bi-layer array, however, for mixed RH-LH stack-structure we assume only different thicknesses of basic slabs, $d_a\neq d_b$. In this case the Bloch phase has finite value, $\gamma\neq0$, and the localization length can be derived in the standard way already used in the previous studies \cite{IK99,KI99,IKU01,HIT08,IM09,IMT10}. Specifically, we expand the coefficients \eqref{FNT-ABCDn} up to the second order in the perturbation parameters $\eta_{a,b}(n)\ll1$. In doing so, one can expand only the factors containing the impedances $Z_{a,b}(n)$. As to the random phase shifts $\widetilde{\varphi}_{a,b}(n)$, they can be replaced with their unperturbed values $\varphi_{a,b}$, see definitions \eqref{FNT-phi-ab}. This fact becomes clear if we take into account the conclusion of Ref.~\onlinecite{IM09}: The phase disorder contributes to the Lyapunov exponent only when the unperturbed impedances are different. The quite cumbersome calculations allow us to derive the perturbed map for the radius $R_n$ and angle $\theta_n$,
\begin{subequations}\label{FNT-mapWD}
\begin{eqnarray}
\frac{R_{n+1}^2}{R_n^2}=1+\eta_{a}(n)V_{a}(\theta_n)+\eta_{b}(n)V_{b}(\theta_n)\nonumber\\[6pt]
+\eta_{a}^2(n)W_{a}+ \eta_{b}^2(n)W_{b}+\eta_{a}(n)\eta_{b}(n)W_{ab},\label{FNT-mapR-WD}\\[6pt]
\theta_{n+1}-\theta_n+\gamma=\eta_{a}(n)U_{a}(\theta_n)+\eta_{b}(n)U_{b}(\theta_n).\label{FNT-mapTheta-WD}
\end{eqnarray}
\end{subequations}
Here the functions standing at random variables $\eta_{a,b}(n)$ are described by the expressions:
\begin{subequations}\label{FNT-VWU}
\begin{eqnarray}
&&V_{a}(\theta_n)=-2\sin\varphi_{a}\sin(2\theta_n-\varphi_{a}),\\
&&V_{b}(\theta_n)=-2\sin\varphi_{b}\sin(2\theta_n-\gamma-\varphi_{a}),\\
&&W_{a}=2\sin^2\varphi_{a},\quad W_{b}=2\sin^2\varphi_{b},\\
&&W_{ab}=4\sin\varphi_{a}\sin\varphi_{b}\cos\gamma;\\
&&U_{a}(\theta_n)=-\sin\varphi_{a}\cos(2\theta_n-\varphi_{a}),\\
&&U_{b}(\theta_n)=-\sin\varphi_{b}\cos(2\theta_n-\gamma-\varphi_{a}).
\end{eqnarray}
\end{subequations}
In Eqs.~\eqref{FNT-VWU} we keep only those terms that contribute to the localization length $L_{loc}$ in the first non-vanishing order of approximation. Note that in Eq.~\eqref{FNT-mapR-WD} the factors $V_{a,b}$ containing $\theta_n$ are always multiplied by $\eta_{a,b}(n)$, therefore, only linear terms in the perturbation are needed in the recurrent relation \eqref{FNT-mapTheta-WD} for the angle $\theta_n$.

Now, in order to evaluate the Lyapunov exponents one has to substitute Eq.~\eqref{FNT-mapR-WD} into Eq.~\eqref{FNT-LyapDef} and expand the logarithm within the quadratic approximation in the perturbation parameters $\eta_{a,b}(n)$. In this approximation one can treat the terms $\eta_{a}^2(n)$, $\eta_{b}^2(n)$ and $\eta_{a}(n)\eta_{b}(n)$ as uncorrelated with factors \eqref{FNT-VWU} containing the angle variable $\theta_n$. It is important that performing the averaging we assume the distribution of phase $\theta_n$ to be homogeneous, i.e., the corresponding distribution function $\rho(\theta)=1/2\pi$. One can show that this assumption is valid apart from the case $\gamma=0$, i.e. when we consider the ideal mixed RH-LH stack-structure with $d_a=d_b$. After some algebra we arrive at the final expression for the Lyapunov exponent,
\begin{eqnarray}\label{FNT-LyapFin}
\lambda=\frac{d}{L_{loc}}&=&\frac{1}{2}\sigma_{a}^2\mathcal{K}_a(2\gamma)\sin^2\varphi_{a}+ \frac{1}{2}\sigma_{b}^2\mathcal{K}_b(2\gamma)\sin^2\varphi_{b}\nonumber\\[6pt]
&&+\sigma_{ab}^2\mathcal{K}_{ab}(2\gamma)\sin\varphi_{a}\sin\varphi_{b}\cos\gamma.
\end{eqnarray}
As one can expect, the result \eqref{FNT-LyapFin} is symmetric with respect to the permutation of slab indices $a\leftrightarrow b$.

The expression \eqref{FNT-LyapFin} for the Lyapunov exponent $\lambda$ consists of three terms. The first two terms contain the \emph{auto-correlators} $\mathcal{K}_a(2\gamma)$ or $\mathcal{K}_b(2\gamma)$, they are originated from the correlations between solely $a$ or solely $b$ slabs, respectively. The third term depends on the \emph{cross-correlator} $\mathcal{K}_{ab}(2\gamma)$ that emerges due to cross-correlations between two disorders $a$ and $b$.

Eq.~\eqref{FNT-LyapFin} is universal and applicable for both homogeneous RH-RH and mixed RH-LH stack-structures. The only difference between these cases is the sign in the unperturbed phase shift $\varphi_{b}=\pm\omega d_b/c$. This affects the value \eqref{FNT-gamma-def} of the Bloch phase $\gamma$ and changes the sign at the third cross-correlation term.

For both homogeneous RH-RH and mixed RH-LH stack-structures, the Lyapunov exponent typically obeys the conventional frequency dependence,
\begin{equation}\label{FNT-LyapOmega2}
\lambda=d/L_{loc}\propto\omega^2\qquad \mathrm{when}\quad\omega\to0.
\end{equation}
However, specific correlations in the disorders of the refractive indices taken into account in Eq.~\eqref{FNT-LyapFin}, may result in a quite unusual $\omega$-dependence. In this respect, of special interest are long-range correlations leading to significant decrease or increase of the localization length $L_{loc}(\omega)$ in the predefined frequency window. Due to these correlations one can enhance or suppress the localization in the systems with compositional disorder (see, e.g., Refs.~\onlinecite{KIKS00,KIK08}).

The expression for the Lyapunov exponent $\lambda(\omega)$ manifests en emergence of the \emph{Fabry-Perot resonances} associated with multiple reflections inside $a$ or $b$ slabs from the interfaces. These resonances occur when the wave frequency $\omega$ meets the conditions,
\begin{equation}\label{FNT-FPab}
\omega/c=s_a\pi/d_a\quad\mathrm{or}\quad\omega/c=s_b\pi/d_b,\quad s_{a,b}=1,2,3,\dots.
\end{equation}
At the resonances the factor $\sin\varphi_a$ or $\sin\varphi_b$ in Eq.~\eqref{FNT-LyapFin} vanishes, resulting in the resonance increase of the localization length $L_{loc}$. When only one type of the basic layers is disordered i.e., Eq.~\eqref{FNT-LyapFin} contains only one corresponding term, the resonances give rise to the divergence of $L_{loc}(\omega)$. In the special case when the ratio between $d_a$ and $d_b$ is a rational number, $d_a/d_b=s_a/s_b$, some resonances from different types of layers coincide. This also leads to the divergence of the localization length. The unexpected feature of these resonances is that they are quite broad due to vanishing of smooth trigonometric functions. As was shown in Refs.~\onlinecite{LIMKS09,LM09,MIL07,IM09}, in the case of positional disorder the terms entering the Lyapunov exponent and related to auto-correlations between the same type of slabs display the Fabry-Perot resonances associated with the other type of layers. On the contrary, in the case of compositional disorder the corresponding terms in Eq.~\eqref{FNT-LyapFin} manifest the Fabry-Perot resonances belonging to the same type of slabs.

\begin{figure}[h]
\begin{center}
\includegraphics[scale=0.79]{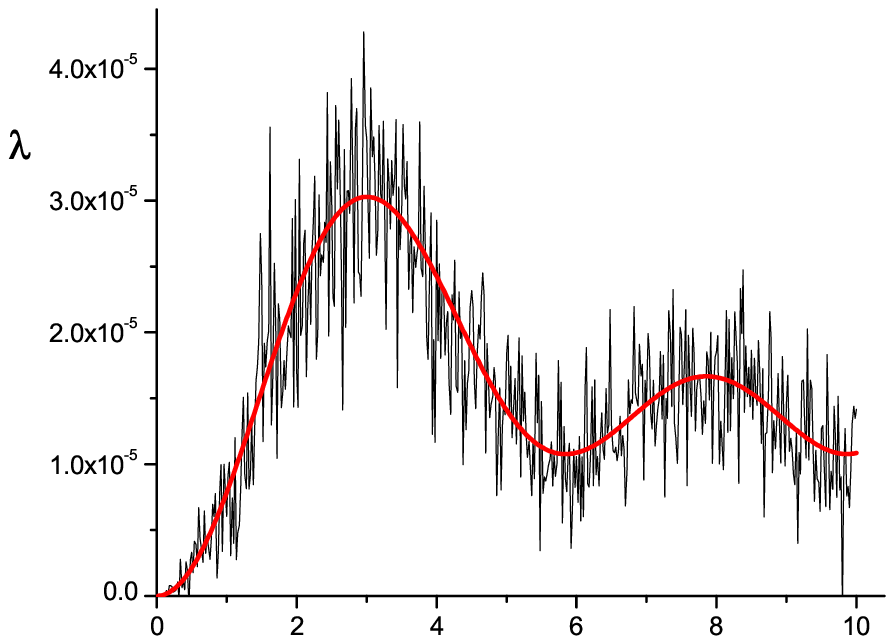}
\includegraphics[scale=0.79]{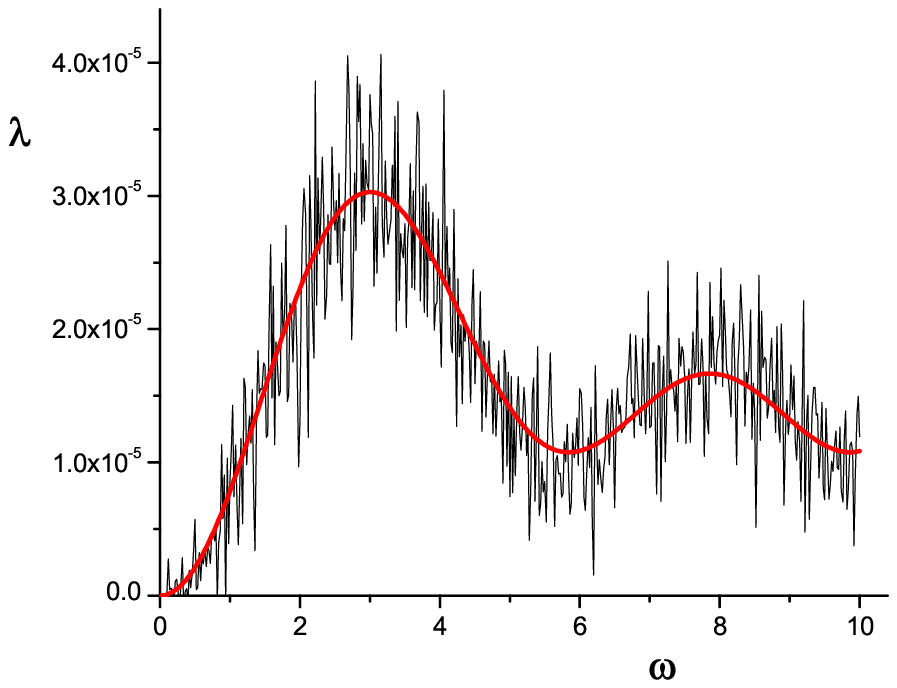}
\caption{\label{MakFNT-Fig0102} (color online) Lyapunov exponent versus frequency. Top: the mixed RH-LH material. Bottom: the RH-RH media. For both cases $\sigma_a\approx\sigma_b\approx 0.006,\, d_a=0.6\,,d_b=0.4\,, c=1$ and the length of sequences is $N=10^6$. Smooth curve corresponds to Eq.(\ref{FNT-LyapWN}).}
\end{center}
\end{figure}

As an example, let us consider the particular case of the white-noise disorders for $a$ and $b$ slabs,
\begin{equation}\label{FNT-CorrFr}
\sigma_{ab}^2=0,\qquad\mathcal{K}_a(k)=\mathcal{K}_b(k)=1.
\end{equation}
In this case the Lyapunov exponent for both homogeneous RH-RH and mixed RH-LH stack-structures takes a quite simple form,
\begin{equation}\label{FNT-LyapWN}
\lambda=\frac{d}{L_{loc}}=\frac{1}{2}(\sigma_a^2\sin^2\varphi_{a}+\sigma_b^2\sin^2\varphi_{b}).
\end{equation}

Numerical results perfectly confirm this dependence, see Fig.~\ref{MakFNT-Fig0102}. The more detailed comparison for $\omega d/c\ll1$ and $\omega d/c\gg1$ also shows a nice correspondence. The data shown here are obtained with the use of Eqs.~\eqref{FNT-mapRTheta} without any approximation. One can see that for a long enough sample and weak disorder the analytical expression \eqref{FNT-LyapWN} perfectly corresponds to the data, apart from random fluctuations. For each case only one realization of the disorder was used, and the fluctuations can be smoothed out by an additional ensemble averaging. In order to see whether our predictions can be applied in experiment, we also used a quite short sample, $N=100$, and strong disorder, $\sigma_a\approx\sigma_b\approx 0.3$. The result shows that the analytical expression is also valid in average, however, the fluctuations around the smooth analytical curve are larger.

\section{Mixed RH-LH stack with vanishing Bloch phase}
\label{vanishingGamma}

As we already noted, the expression \eqref{FNT-LyapFin} for the Lyapunov exponent is valid for {\it both} the homogeneous RH-RH and mixed RH-LH bi-layer array when the Bloch phase \eqref{FNT-gamma-def} is essentially different from zero. In this case the unperturbed map \eqref{FNT-mapRThetaUnp} does not degenerate, and nothing special arises for the evaluation of the Lyapunov exponent. Physically this is related to the fact that the unperturbed phase $\theta_n$ rotates when passing the sample of length $N=L/d$ resulting in a homogeneous randomization. This happens everywhere provided the Bloch phase $\gamma$ are irrational with respect to $2\pi$. Note that in this case the $\theta$-distribution can be regarded as homogeneous even without the disorder.

The situation is completely different in the case when the wave phase $\gamma=\varphi_a+\varphi_b$ vanishes after passing the unit $(a,b)$ cell. This happens in the ideal mixed RH-LH stack with $d_a=d_b$ because after passing the RH $a$-layer, the phase shift is $\varphi_{a}=\omega d_a/c$, however, it is exactly canceled by another shift, $\varphi_{b}=-\omega d_a/c=-\varphi_a$ in the next LH $b$-slab. As one can see, the circle \eqref{FNT-mapRThetaUnp} presenting the unperturbed map, degenerates into a single point in the phase space $(Q,P)$, and, therefore, the unperturbed phase distribution is simply delta-function. This means that with a weak disorder, the phase distribution should not be expected as homogeneous one.

In what follows, for simplicity we consider the ideal mixed stack whose refractive indices, $n_a$ and $n_b$, are perturbed by two independent white-noise disorders with the same strength \cite{Ao07,Ao10},
\begin{subequations}\label{FNT-FreWN-IMS}
\begin{eqnarray}
\sigma_{a}^2=\sigma_{b}^2\equiv\sigma^2,\quad\sigma_{ab}^2=0,\quad\mathcal{K}_a(k)=\mathcal{K}_b(k)=1;\quad\label{FNT-FreWN}\\[6pt]
d_a=d_b,\quad\mathrm{i.e.,}\quad\varphi_{a}=-\varphi_b\equiv\varphi\quad\mathrm{and}\quad\gamma=0.\quad\label{FNT-IMS}
\end{eqnarray}
\end{subequations}

The phase distribution $\rho(\theta)$ can be found in the similar way as was described in Refs.~\onlinecite{IKT95,HIT08}. The starting point is the exact recurrent relation \eqref{FNT-mapTheta} between $\theta_{n+1}$ and $\theta_n$. By expanding this expression up to the second order in perturbation parameters $\eta_{a,b}(n)$, one obtains
\begin{equation}\label{FNT-mapTheta-Fre}
\theta_{n+1}-\theta_{n}=-[\eta_a(n)-\eta_b(n)]v(\theta_n)+\sigma^2v(\theta_n)v'(\theta_n),
\end{equation}
where we introduced the function
\begin{equation}\label{FNT-Vdef}
v(\theta)=\varphi+\sin\varphi\cos(2\theta-\varphi).
\end{equation}
In deriving Eq.~\eqref{FNT-mapTheta-Fre}, we kept the linear terms proportional to $\eta_a(n)$, $\eta_b(n)$ and substituted the terms
$\eta_a^2(n)$, $\eta_b^2(n)$ by $\langle\eta_a^2(n)\rangle=\langle\eta_b^2(n)\rangle=\sigma^2$. Also, we explicitly took into account the condition $\langle\eta_a(n)\eta_b(n)\rangle=0$ that directly follows from Eqs.~\eqref{FNT-CorrDef} and \eqref{FNT-FreWN}. This approximation is sufficient in order to obtain the distribution of phases $\theta_n$ in the second order of perturbation. As a result, the expression (\ref{FNT-mapTheta-Fre}) takes the form of the stochastic Ito equation \cite{G04} which can be associated with the Fokker-Plank equation for the distribution function $P(\theta, t)$ (see also Ref.~\cite{IRT98}),

The next step is to obtain the differential equation for the probability density $\rho(\theta)$. In the case of weak disorder the corresponding Fokker-Plank equation for the distribution function $P(\theta, t)$ has a relatively simple form,
\begin{equation}\label{FNT-FPeq}
\frac{\partial^2P}{\partial t^2}=\sigma^2\frac{\partial^2}{\partial\theta^2}\left[P(\theta,t)v^2(\theta)\right]- \frac{\sigma^2}{2}\frac{\partial}{\partial\theta}\left[P(\theta,t)\frac{dv^2(\theta)}{d\theta}\right].
\end{equation}
Here the "time" $t$ is the same as the length $N$ of a sample along which the evolution of phase $\theta$ occurs.

Since we are interested in the stationary solution of this equation, $\rho(\theta)\equiv P(\theta,t\to\infty)$, the equation for $\rho(\theta)$ reads
\begin{equation}\label{FNT-rho-eq}
\frac{d^2}{d\theta^2}\left[\rho(\theta)v^2(\theta)\right]-\frac{1}{2}\frac{d}{d \theta}\left[\rho(\theta)\frac{d}{d \theta}v^2(\theta)\right]=0.
\end{equation}
Here the dependence on the variance $\sigma^2$ has disappeared due to the rescaling of time, $t\to\sigma^2t$. Therefore, in this approximation the phase probability density $\rho(\theta)$ does not depend on the disorder variance $\sigma^2$. Note that the only function $v(\theta)$ entering the diffusive equation \eqref{FNT-rho-eq}, is periodic with the period $\pi$. Consequently, its solution $\rho(\theta)$ should be also periodic with the same period. In addition, $\rho(\theta)$ should satisfy the normalization condition,
\begin{equation}\label{FNT-rho-Cond}
\rho(\theta+\pi)=\rho(\theta),\qquad\qquad\int_{0}^{\pi}d\theta\rho(\theta)=1.
\end{equation}
One can easily obtain that the solution of Eqs.~(\ref{FNT-rho-eq}) and \eqref{FNT-rho-Cond} is
\begin{equation}\label{FNT-rho-sol}
\rho(\theta)=J/v(\theta)\,,\qquad J=\frac{1}{\pi}\sqrt{\varphi^2 -\sin^2\varphi}\,.
\end{equation}

\begin{figure}[h!!!]
\begin{center}
\includegraphics[scale=0.30]{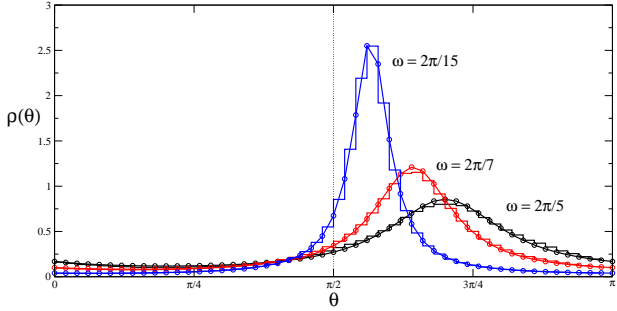}
\end{center}
\caption{(color online) Stationary distribution $\rho(\theta)$ for various values of $\omega=\varphi$ in rescaled units $d_a=1$, $c=1$. Broken curves correspond to numerical data with an ensemble average for $N=10^6$, $10^7$, $10^8$ for $\omega=2\pi/5, 2\pi/7,2\pi/15$, respectively. Smooth curves present the analytical expression \eqref{FNT-rho-sol}.}\label{FNT-Fig03}
\end{figure}

Our results indicate that the phase distribution $\rho(\theta)$ strongly depends on the phase shift $\varphi$, and is highly inhomogeneous in the limit $\varphi\equiv\varphi_{a}=\omega d_a/c\ll1$, i.e., for small values of the wave frequency $\omega$. Some examples of the distribution function $\rho(\theta)$ for different values of $\varphi$ are shown in Fig.~\ref{FNT-Fig03}. This figure clearly demonstrates that with a decrease of $\omega$ the distribution $\rho (\theta)$ starts to be very sharp in the vicinity of $\theta = \pi/2$.

It is worthwhile to note that the situation for the vanishing value of $\gamma$ is somewhat similar to that emerging for the Anderson and Kronig-Penney models at the band edges. Indeed, in these models the unperturbed Bloch phase $\gamma$ also vanishes when approaching the band edges. This leads to a highly non-homogeneous distribution of perturbed phase, and, as a result, to a non-standard dependence of the Lyapunov exponent on the model parameters.

\begin{figure}[h!]
\vspace{1cm}
\begin{center}
\includegraphics[scale=0.55]{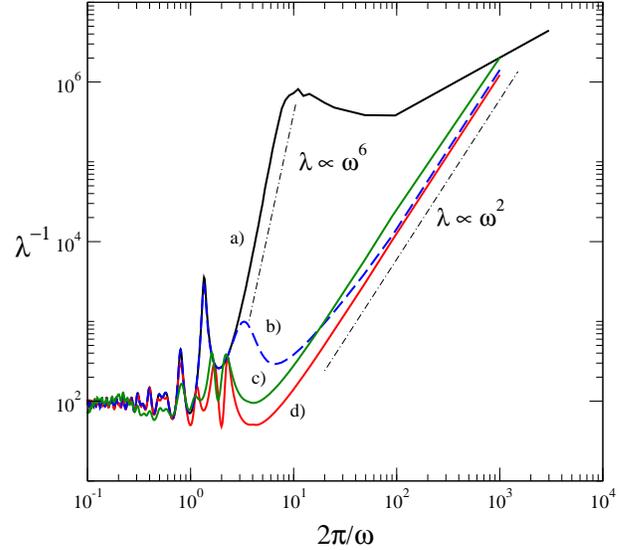}
\end{center}
\caption{(color online) Inverse Lyapunov exponent versus the rescaled frequency $\omega$ with the variance of disorder $\sigma^2=0.02$ and sequence length $N=10^6$. The data are obtained with an ensemble averaging over 100 realizations of disorder; a) mixed RH-LH array with $d_a=d_b$, b) the same for $d_a=0.99d_b$, c) the same for $d_a=0.1d_b$, d) homogeneous RH-RH array with $d_a=d_b$. For comparison, the dot-dashed lines show two frequencies dependencies discussed in the text.}\label{FNT-Fig04}
\end{figure}

In the considered model of the mixed RH-LH array, the crucial difference is that the effect related to the value $\gamma=0$ emerges {\it independently of the frequency} $\omega$. This is in contrast with the case of Anderson and Kronig-Penney models for which the zero Bloch phase occurs at band edges only, therefore, for specific values of frequency. Thus, one can expect that for mixed RH-LH bi-layer stacks with the specific condition $d_a=d_b$, the dependence of the Lyapunov exponent on the model parameters has to be highly non-trivial. This fact can be seen from Fig.~\ref{FNT-Fig04} which shows how the dependence of the Lyapunov exponent on frequency $\omega$ is affected by the relation between $d_a$ and $d_b$.

The data in this figure clearly display that when $d_a$ approaches $d_b$, the standard dependence $\lambda\propto\omega^2$ that is known to be a generic case for small $\omega$, alternates by a very unusual dependence $\lambda\propto\omega^6$. The latter result was found numerically in Ref.~\onlinecite{Ao07} and later on, was discussed in Refs.~\onlinecite{Ao10,Mo10}. From Fig.~\ref{FNT-Fig04} one can also see that for the homogeneous RH-RH stack-structure the conventional dependence \eqref{FNT-LyapOmega2} remains valid even in the specific case of $d_a=d_b$. As was shown above, in this case the Lyapunov exponent $\lambda$ is described by the generic expression \eqref{FNT-LyapFin} that for homogeneous RH-RH array is valid for any relation between $d_a$ and $d_b$.

The Lyapunov exponent can be derived according to definition \eqref{FNT-LyapDef} and exact Hamiltonian map \eqref{FNT-mapR} for the radius $R_n$. Its expansion within the second order of approximation in the disorder $\eta_{a,b}(n)$ gives
\begin{equation}\label{FNT-LyapRHLH}
\lambda=2\sigma^2\sin\varphi\langle\cos(2\theta_n-\varphi)v(\theta_n)\rangle\,.
\end{equation}
The averaging in this expression has to be performed with the distribution function $\rho(\theta)$ determined by Eqs.~\eqref{FNT-rho-sol} and \eqref{FNT-Vdef}. Since the denominator $v(\theta)$ in Eq.~\eqref{FNT-rho-sol} is the same as the coefficient in Eq.~\eqref{FNT-LyapRHLH}, we come to the remarkable result that the Lyapunov exponent \eqref{FNT-LyapRHLH} vanishes for any value of the phase shift $\varphi$ ! This means that in order to derive the non-vanishing Lyapunov exponent, one has to obtain the expressions for both the phase distribution $\rho(\theta)$ and the ratio $R^2_{n+1}/R^2_n$ in the next order of perturbation, by expanding them up to the fourth order in disorder which is not a simple task. Thus, one can expect that the Lyapunov exponent for the ideal mixed RH-LH stack with $d_a=d_b$, should be proportional to $\sigma^4$ in contrast with the conventional quadratic dependence, $\lambda\propto\sigma^2$. Our extensive numerical and analytical studies confirm this expectation.

\section{acknowledgments}
We dedicate this paper to the memory of Prof. E.A. Kaner on the occasion of his 80th birthday and 55th anniversary of prediction of the Azbel-Kaner resonance.

F.M.I acknowledges support from VIEP grant EXC08-G of the BUAP (Mexico).




\begin{thebibliography}{99}

\bibitem{MS08}
P. Marko\v{s}, C.M. Soukoulis, \emph{Wave Propagation. From Electrons to Photonic Crystals and Left-Handed Materials}, Princeton University Press, Princeton (2008).

\bibitem{IK99}
F.M. Izrailev, A. Krokhin, \emph{Phys. Rev. Lett.} \textbf{82}, 4062 (1999).

\bibitem{KI99}
A.A. Krokhin, F.M. Izrailev, \emph{Ann. Phys. (Leipzig)} \textbf{8}, 153 (1999).

\bibitem{IKU01}
F.M. Izrailev, A.A. Krokhin, S.E. Ulloa, \emph{Phys. Rev. B} \textbf{63}, 041102(R) (2001).

\bibitem{IM01}
F.M. Izrailev, N.M. Makarov, \emph{Opt. Lett.} \textbf{26}, 1604-1106 (2001).

\bibitem{IM03}
F.M. Izrailev, N.M. Makarov, \emph{Phys. Rev. B} \textbf{67}, 113402 (2003).

\bibitem{IM04} F.M. Izrailev, N.M. Makarov, \emph{Appl. Phys. Lett.} \textbf{84}, 5150-5152 (2004).

\bibitem{IM05}
F.M. Izrailev, N.M. Makarov, \emph{J. Phys. A: Math. Gen.} \textbf{38}, 10613 (2005).

\bibitem{KIKS00}
U. Kuhl, F.M. Izrailev, A.A. Krokhin, H.-J. St\"ockmann, \emph{Appl. Phys. Lett.} \textbf{77}, 633 (2000).

\bibitem{KIK08}
U. Kuhl, F.M. Izrailev, A.A. Krokhin, \emph{Phys. Rev. Lett.} \textbf{100}, 126402 (2008).

\bibitem{DKSMI11}
O. Dietz, U. Kuhl, H.-J. St\"{o}ckmann, N.M. Makarov, F.M. Izrailev, \emph{Phys. Rev. B} \textbf{83}, 134203 (2011).

\bibitem{LGP88}
I.M. Lifshits, S.A. Gredeskul, L.A. Pastur, \emph{Introduction to the Theory of Disordered Systems}, Wiley, New York (1988).

\bibitem{M99}
N.M. Makarov, Lectures on \emph{Spectral and Transport Properties of One-Dimensional Disordered Conductors} (1999), $(http://www.ifuap.buap.mx/virtual/page_vir.html)$.

\bibitem{MCMM93}
A.R. McGurn, K.T. Christensen, F.M. Mueller, A.A. Maradudin, \emph{Phys. Rev. B} \textbf{47}, 13120 (1993).

\bibitem{So00}
D.R. Smith, W.J. Padilla, D.C. Vier, S.C. Nemat-Nasser, S. Schultz, \emph{Phys. Rev. Lett.} \textbf{84}, 4184 (2000).

\bibitem{SSS01}
R.A. Shelby, D.R. Smith, S. Schultz, \emph{Science} \textbf{292}, 77 (2001).

\bibitem{Po03}
C.G. Parazzoli, R. B. Greegor, K. Li, B. E. C. Koltenbah, M. Tanielian, \emph{Phys. Rev. Lett.} \textbf{90}, 107401 (2003).

\bibitem{VM03}
A.P. Vinogradov, A.M. Merzlikin, \emph{Physica B} \textbf{338}, 126 (2003).

\bibitem{Eo06}
A. Esmialpour, M. Esmaeilzadeh, E. Faizabadi, P. Carpena, M.R.R. Tabar, \emph{Phys. Rev. B} \textbf{74}, 024206 (2006).

\bibitem{DZ06}
Yu. Dong, X. Zhang, \emph{Phys. Lett. A} \textbf{359}, 542 (2006).

\bibitem{No07}
D. Nau, A. Schoenhardt, Ch. Bauer, A. Christ, T. Zentgraf, J. Kuhl, M.W. Klein, H. Giessen, \emph{Phys. Rev. Lett.} \textbf{98}, 133902 (2007).

\bibitem{Po07}
I.V. Ponomarev, M. Schwab, G. Dasbach, M. Bayer, T.L. Reinecke, J.P. Reithmaier, A. Forchel,  \emph{Phys. Rev. B} \textbf{75}, 205434 (2007).

\bibitem{Ao07}
A.A. Asatryan, L.C. Botten, M.A. Byrne, V.D. Freilikher, S.A. Gredeskul, I.V. Shadrivov, R.C. McPhedran, Yu.S.Kivchar, \emph{Phys. Rev. Lett.} \textbf{99}, 193902 (2007).

\bibitem{NML07}
E.M. Nascimento, F.A.B.F. de Moura, M.L. Lyra, \emph{Phys. Rev. B} \textbf{76}, 115120 (2007).

\bibitem{NML08}
E.M. Nascimento, F.A.B.F. de Moura, M.L. Lyra, \emph{Optics Express} \textbf{16}, 6860 (2008).

\bibitem{BW85}
V. Baluni, J. Willemsen, \emph{Phys. Rev. A} \textbf{31}, 3358 (1985).

\bibitem{LIMKS09}
G.A. Luna-Acosta, F.M. Izrailev, N.M. Makarov, U. Kuhl, H.-J. St\"{o}ckmann, \emph{Phys. Rev. B} \textbf{80}, 115112 (2009).

\bibitem{LM09}
G.A. Luna-Acosta, N.M. Makarov, \emph{Ann. Phys. (Berlin)} \textbf{18}, 887 (2009).

\bibitem{Ao10}
A.A. Asatryan, S.A. Gredeskul, L.C. Botten, M.A. Byrne, V.D. Freilikher,  I.V. Shadrivov, R.C. McPhedran, Yu.S.Kivchar, \emph{Phys. Rev. B} \textbf{81}, 075124 (2010).

\bibitem{Mo10}
D. Mogilevtsev, F.A. Pinheiro, R.R. dos Santos, S.B. Cavalcanti, L.E. Oliveira, \emph{Phys. Rev. B} \textbf{82}, 081105(R) (2010).

\bibitem{HIT08}
J.C. Hern\'{a}ndez-Herrej\'{o}n, F.M. Izrailev, L. Tessieri, \emph{Physica E} \textbf{40}, 3137 (2008).

\bibitem{MIL07}
N.M. Makarov, F.M. Izrailev, G. Luna-Acosta, in: \emph{Proc. 6th Int. Kharkov Symposium on Physics and Engineering of Microwaves, Millimeter, and Submillimeter Waves and Workshop on Terahertz Technologies} (Kharkov, 2007) vol~1, p~140-145 (IEEE Catalog Number: 07EX1786, Library of Congress: 2007925116).

\bibitem{IM09}
F.M. Izrailev, N.M. Makarov, \emph{Phys. Rev. Lett.} \textbf{102}, 203901 (2009).

\bibitem{IMT10}
F.M. Izrailev, N.M. Makarov, E.J. Torres-Herrera, \emph{Physica B: Condensed Matter} \textbf{405}, 3022 (2010).

\bibitem{MIT10}
N.M. Makarov, F.M. Izrailev, E.J. Torres-Herrera, in: \emph{2010 International Kharkov Symposium on Physics and Engineering of Microwaves, Millimeter and Submillimeter Waves. Sessions and papers} (Kharkov, 2010) 5~pp (IEEE Catalog Number: CFP10780-CDR).

\bibitem{IKT95}
F.M. Izrailev, T. Kottos, G. Tsironis, \emph{Phys. Rev. B} \textbf{52}, 3274 (1995).

\bibitem{IRT98}
F.M. Izrailev, S. Ruffo, L. Tessieri,  \emph{J. Phys. A: Math. Gen.} \textbf{31}, 5263 (1998).

\bibitem{G04}
C.W. Gardiner, \emph{Handbook of Stochastic Methods}, 3nd ed., Springer-Verlag, Berlin (2004).

\end{thebibliography}
\end{document}